# Turbulence Calculation Based on the Extended Navier–Stokes Equations

Shanwen Tan [1*], Zhengui Li [2*], Wangxu Li [3]

[1] School of Architecture and Civil Engineering, Xihua University, Chengdu, Sichuan, 610039

[2] Key Laboratory of Fluid and Power Machinery of the Ministry of Education, Xihua University, Chengdu, Sichuan, 610039

[3] School of Energy and Power Engineering, Lanzhou University of Technology, Lanzhou, Gansu, 730050

(1.email: tanshanwen68@mail.xhu.edu.cn)

**Abstract:** In the study, we proposed a computational method for solving the turbulence problem of incompressible viscous Newtonian fluids based on extended Navier–Stokes (N–S) equations. With several phenomenological observations and H. J. Kreuer's interpretation of the origin of viscosity, we make a hypothesis that the shear process in fluid flow is accompanied by eddy formation, and considering the influence of eddies on convection and diffusion, the classical N–S equations have been improved to obtain the extended N–S equations. The extended equations are closed, and the source of the velocity fluctuations being explicitly considered to be additional convection and diffusion. The extended equations are compatible with the classical N–S equations and can describe laminar and turbulent flows in a unified manner. In fluid flow simulations, the equations describing the mean flow quantities can be obtained directly from the extended N–S equations without any additional turbulence models. The flow over a cube placed in a channel was numerically investigated to verify the extended equation, the simulation results agreeing well with the Large Eddy Simulation (LES) and experimental results.

## 1. Introduction

Turbulent flows are widespread in engineering and nature. In fluid mechanics, it is believed that the Navier–Stokes (N–S) equations—supplemented by appropriate boundary and initial conditions—are sufficient to describe turbulent flows. To date, researchers have not found a general analytical solution to the N–S equations, and for most engineering applications, we can only resort to numerical methods to obtain results [1–2].

Direct numerical simulation (DNS) is a direct approach for solving the N–S equations. In turbulent flow, the flow behavior is random and chaotic and spans a wide range of time and length scales, making the DNS of turbulent flow difficult.

Owing to certain statistical properties of turbulence, research and engineering applications have focused on the mean values of hydrodynamic quantities to describe turbulent flows. Currently, two categories of averaging methods are used in mainstream turbulence simulations—that is, the Reynolds-averaged Navier–Stokes (RANS) equations based on time averaging, and Large Eddy Simulation (LES) based on space averaging. In the RANS equations, the mean flow equations are obtained by using time averaging to the N–S equations, owing to the interactions between different fluctuating components, extra terms appearing in the time-averaged N–S equations, these extra terms can be approximated by using the appropriate turbulence models to close the RANS equations. The LES uses a space filter to the N–S equations, the filter passing the larger eddies and rejecting the smaller ones, the mean flow equations governing the evolution of the mean flow and the large eddies, and the unresolved small-scale eddies being parameterized using a subgrid model. For the RANS equations and LES, the closure of the equations depends on the appropriate turbulence model, all of which have been developed based on semi-empirical and dimensional analysis methods, which can be valuable for solving a number of important practical problems. However, the hypotheses used in these models have no reliable physical basis and contribute little to our understanding of the physical nature of turbulence [3–8].

To date, there are no universal turbulence closure models, mainly because the mean flow equations cannot be directly obtained from the N–S equations in turbulence simulations [9]. This means that the N–S equations are not sufficient to describe turbulence. Moreover, this is still an open problem in three dimensions, and it is generally accepted that three-dimensional N–S equations are not a complete description of fluid flows [10].

In this study, we proposed ideas aimed at improving the existing equations of fluid motion. Based on several phenomenological observations — in particular, Kreuer's interpretation of the origin of viscosity—we assumed that the shear process is accompanied by eddy formation, considering the influence of eddies on convection and diffusion, the classical N–S equations were improved to obtain extended N–S equations. The extended N–S equations explicitly included the source that produces the fluctuating components, thus, in the turbulence simulation, we could obtain equations describing the average quantities directly from the extended equations without any additional turbulence model, furthermore the extended equations unified the simulation of laminar and turbulent flows. The flow over a cube placed in a channel was numerically investigated to verify the extended equation, the simulation results agreeing well with the LES and experimental results.

## 2. Methods: Setup of the extended N–S equations

### 2.1 Establishment of the model

In continuum mechanics—including solid mechanics and fluid mechanics—there are two fundamental dynamic processes: the compressive/expand ("compressive" in short) process and the shear process. These two fundamental processes, and their coupling, play important roles in continuum mechanics. In the compressive process, a solid and fluid have the same response to normal stress—for example, neither a solid nor a fluid can be compressed infinitely. In the shear process, the response of a solid and a fluid to shear stress are completely different, where a solid can remain in equilibrium with finite deformation, but fluid flow will last forever.

To describe the fluid flow, we can imagine that the fluid comprises many layers that can slide over one another. Under the action of the shear process, the flow layers will slide at different speeds, an ideal fluid has no viscosity, the shear process can force one layer of fluid to slide over another without friction, causing a tangential discontinuity in the velocity. In reality, all fluids have more or less shear viscosity, and shear stress always causes the fluid elements between two adjacent fluid layers to rotate, creating some vortex structures. Moreover, as the Reynolds number increases, the vortices can further evolve, interact, become unstable, and break into turbulence, making the fluid flow extremely complex. These phenomena never occur during solid motion. Consequently, the late Prof. Shi-Jia Lu (1911–1986) — the only female student of Ludwig Prandtl — made a very insightful assertion around 1980: "The essence of a fluid is vortices. A fluid cannot stand rubbing; once you rub it, vortices appear" [11].

Next, we discuss the source of fluid viscosity. When a velocity field with a velocity gradient acts on a fluid, a resistance is created within the fluid to weaken such inhomogeneities. This can be explained by the friction between the fluid layers —that is, the viscosity of the fluid. According to Kreuer's interpretation, the source of the fluid viscosity come from two physical effects[12]. The first physical effect that dominates in gases is that of collision. At a microscopic scale within fluid flow, if two adjacent fluid layers move at different velocities, the molecules in the two fluid layers tend to exchange and collide with each other as a result of thermal excitation. This causes the faster fluid layer to decelerate and the slower fluid layer to accelerate, weakening the velocity discrepancy between them. This physical effect is generally accepted by the hydrodynamic community to be a source of viscosity; however, this interpretation is not applicable to liquids.

The second physical effect—which reduces the inhomogeneity of fluid motion—comes from the intermolecular forces that are dominant in liquids. The fluid molecules adjacent to each other are always confined to the range of their mutual interactions; consequently, their mean free path is close to zero, and the viscous effects caused by collisions are negligible. However, when two adjacent layers slide at different speeds, the molecules at their contact surfaces stick together because of intermolecular forces, creating a secondary viscous effect. Moreover, such stickiness between molecules can produce self-organized structures—such as eddies — which cannot be ignored at a macroscopic scale. For liquids, this effect seems to better explain the source of viscosity.

Having presented these phenomenological observations and Kreuer's interpretation of the origin of viscosity, we can hypothesize that in fluid flow, the shear process is accompanied by vortex formation—that is, each differential element is attached to an eddy, and the center of the eddy is the same as the center of the element.

According to Kolmogorov's scale theory, there exists a smallest scale of eddy motion in turbulent flow, and eddies living on scales smaller than this are dissipated into internal thermal energy because of viscosity. The eddy size at the smallest scale is proportional to $v^{3/4}$, where $v$ denotes the kinematic viscosity of the fluid [13–14]. The eddy attached to the differential element is the same size as that of the eddy at the smallest scale.

Fluid flow is formed by the combined action of the external conditions, shearing, compression, and rotating vortices. To derive the governing equations of the fluid flow, we can implement them in two steps. Here, we consider only incompressible, viscous, and Newtonian fluids.

**Step 1**: If only compressive and shear processes occurred in the fluid flow, we can get the classical Navier–Stokes equations [7–8] and be expressed as follows:

$$\nabla \boldsymbol{u} = 0 \qquad (1)$$

$$\rho\left(\frac{\partial \boldsymbol{u}}{\partial t} + (\boldsymbol{u}\nabla)\boldsymbol{u}\right) = -\nabla p + \rho \boldsymbol{g} + \mu \nabla^2 \boldsymbol{u} \qquad (2)$$

where $t$ denotes the time, $\boldsymbol{u}$ and $p$ denote the velocity vector and pressure of the flow field, respectively, and $\mu$ denotes the dynamic viscosity of the fluid.

**Step 2**: Add the effects of the eddy to the fluid flow.

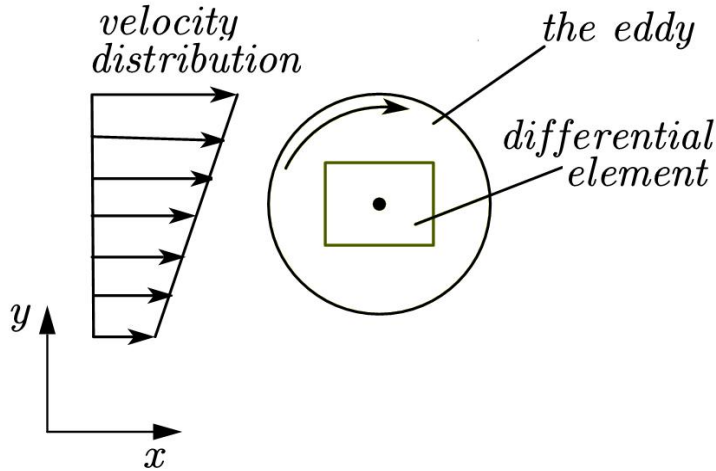

**Figure 1:** The differential element and attached eddy in the field of flow with a velocity gradient.

We can now discuss the effect of the eddy on fluid flow. As shown in **Figure 1**, the rotational motion of the eddy does not change its mass distribution but carries momentum into and out of the differential element. Because of the presence of the velocity gradient, the momentum entering and leaving the differential element is generally not equal—that is, the rotation of the eddy produces an additional exchange of momentum in the differential element, and the velocity distribution of the differential element changes. Consequently, the momentum equations should be affected by the presence of an eddy.

There are two basic processes in fluid flow—that is, convection and diffusion. The effect of eddies can be decomposed into the effects of convection and diffusion and can be added to **Equation (2)**, as follows:

$$\rho\left(\frac{\partial \boldsymbol{u}}{\partial t}+(\boldsymbol{u}\nabla)\boldsymbol{u}+\boldsymbol{R}_1(\boldsymbol{u})\right)=-\nabla p+\rho\boldsymbol{g}+\mu\nabla^2\boldsymbol{u}+\boldsymbol{R}_2(\boldsymbol{u}) \tag{3}$$

where $\boldsymbol{R}_1(\boldsymbol{u})$ denotes the effect of the eddy on convection and $\boldsymbol{R}_2(\boldsymbol{u})$ denotes the effect of the eddy on diffusion.

To demonstrate the completeness of the governing equations, we can use the momentum equation in the *x*-direction as an example, as follows:

$$\rho\left(\frac{\partial u}{\partial t}+u\frac{\partial u}{\partial x}+v\frac{\partial u}{\partial y}+w\frac{\partial u}{\partial z}+R_1(u)\right)=-\frac{\partial p}{\partial x}+\rho g_x+$$
$$\mu\left(\frac{\partial^2 u}{\partial x^2}+\frac{\partial^2 u}{\partial y^2}+\frac{\partial^2 u}{\partial z^2}\right)+R_2(u) \tag{4}$$

where $R_1(u)$ and $R_2(u)$ denote the component in the *x*-direction of $\boldsymbol{R}_1(\boldsymbol{u})$ and $\boldsymbol{R}_2(\boldsymbol{u})$, respectively.

Firstly, the function $R_1(u)$ can be determined. Based on the above hypothesis, the center of the differential element is the same as the center of the eddy, so the convection velocity does not change due to the rotation of the eddy. Consequently, the effect of the eddy on convection ($R_1(u)$) can be further assigned to the convection terms, and **Equation (4)** can be adjusted, as follows:

$$\rho\left(\frac{\partial u}{\partial t}+u\left(\frac{\partial u}{\partial x}+R_{11}(u)\right)+v\left(\frac{\partial u}{\partial y}+R_{12}(u)\right)+w\left(\frac{\partial u}{\partial z}+R_{13}(u)\right)\right)=$$
$$-\frac{\partial p}{\partial x}+\rho g_x+\mu\left(\frac{\partial^2 u}{\partial x^2}+\frac{\partial^2 u}{\partial y^2}+\frac{\partial^2 u}{\partial z^2}\right)+R_2(u) \tag{5}$$

Using the convection component ($R_{12}(u)$) as an example, we can determine the convection correction items. **Figure 2** shows the velocity distribution in the *y*-direction and its change owing to eddy rotation, where $O$ denotes the center of the eddy, $r$ denotes the radius of the eddy, $\widehat{AB}$ denotes the primary velocity distribution, and $\overline{AB}$ denotes the final velocity distribution owing to the eddy rotation.

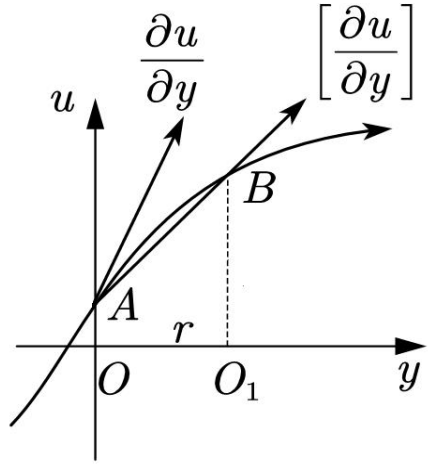

**Figure 2:** The velocity distribution in the *y*-direction and how it is redistributed due to the eddy rotation

The eddy rotation does not destroy the velocity continuity, but the velocity distribution near the center of the eddy changes. The curve $\widehat{AB}$—that is, the primary velocity distribution—changes to $\overline{AB}$ over the entire extent of the eddy, and the first derivative term changes from $\frac{\partial u}{\partial y}$ to $\left[\frac{\partial u}{\partial y}\right]$, so the corresponding convection term can be corrected as follows:

$$v\left[\frac{\partial u}{\partial y}\right] = v\left(\frac{\partial u}{\partial y} + R_{12}(u)\right) \qquad (6)$$

For similar reasons:

$$u\left[\frac{\partial u}{\partial x}\right] = u\left(\frac{\partial u}{\partial x} + R_{11}(u)\right), \quad w\left[\frac{\partial u}{\partial z}\right] = w\left(\frac{\partial u}{\partial z} + R_{13}(u)\right) \qquad (7)$$

Assuming that the velocity at the center of the eddy is $u(O)$, the velocity at point $O_1$ is $u(O_1)$, and using the Taylor series, $u(O_1)$ can be expressed as follows (only the first three items being retained):

$$u(O_1) = u(O) + \frac{\partial u}{\partial y}r + \frac{1}{2!}\frac{\partial^2 u}{\partial y^2}r^2 + \ldots \qquad (8)$$

The ultimate form of the first derivative of the velocity can then be expressed as follows:

$$\left[\frac{\partial u}{\partial y}\right] = \frac{u(O_1) - u(O)}{r} \approx \frac{\partial u}{\partial y} + \frac{1}{2!}\frac{\partial^2 u}{\partial y^2}r \tag{9}$$

The coefficient of the second derivative term can be defined as a new parameter ($\delta$), where $\delta = r/2!$. This is called the scale coefficient, and **Equation (9)** can be rewritten as follows:

$$\left[\frac{\partial u}{\partial y}\right] = \frac{\partial u}{\partial y} + \delta \frac{\partial^2 u}{\partial y^2} \tag{10}$$

Comparing **Equation (10)** with **Equation (6)**, the convection correction term can be expressed as follows:

$$R_{12}(u) = \delta \frac{\partial^2 u}{\partial y^2} \tag{11}$$

Similarly:

$$R_{13}(u) = \delta \frac{\partial^2 u}{\partial z^2} \tag{12}$$

Because the rotation of the eddy has no compression or expansion effect, it has no effect on the radial velocity distribution of the eddy. Consequently:

$$R_{11}(u) = 0 \tag{13}$$

Substituting **Equations (11), (12), (13)** into **Equation (5)**, we can obtain the following:

$$\rho\left(\frac{\partial u}{\partial t} + u\left(\frac{\partial u}{\partial x}\right) + v\left(\frac{\partial u}{\partial y} + \delta \frac{\partial^2 u}{\partial y^2}\right) + w\left(\frac{\partial u}{\partial z} + \delta \frac{\partial^2 u}{\partial z^2}\right)\right) = \\ -\frac{\partial p}{\partial x} + \rho g_x + \mu\left(\frac{\partial^2 u}{\partial x^2} + \frac{\partial^2 u}{\partial y^2} + \frac{\partial^2 u}{\partial z^2}\right) + R_2(u) \tag{14}$$

The expression of $R_2(u)$ can then be determined using the following derivation. To obtain the correction terms for diffusion, the viscous terms of the governing equation can be expressed in stress style. The momentum equation in the *x*-direction can be expressed as follows:

$$\rho\left(\frac{\partial u}{\partial t} + u\frac{\partial u}{\partial x} + v\frac{\partial u}{\partial y} + w\frac{\partial u}{\partial z}\right) = -\frac{\partial p}{\partial x} + \rho g_x + \frac{\partial \tau_{xx}}{\partial x} + \frac{\partial \tau_{yx}}{\partial y} + \frac{\partial \tau_{zx}}{\partial z} \tag{15}$$

For Newtonian fluids, the constitutive equations are as follows:

$$\tau_{xx} = 2\mu\frac{\partial u}{\partial x}, \quad \tau_{yx} = \mu\left(\frac{\partial u}{\partial y} + \frac{\partial v}{\partial x}\right), \quad \tau_{zx} = \mu\left(\frac{\partial u}{\partial z} + \frac{\partial w}{\partial x}\right) \tag{16}$$

Substituting **Equation (16)** into **Equation (15)**, we get the following:

$$\rho\left(\frac{\partial u}{\partial t}+u\frac{\partial u}{\partial x}+v\frac{\partial u}{\partial y}+w\frac{\partial u}{\partial z}\right)=-\frac{\partial p}{\partial x}+\rho g_x+$$
$$\frac{\partial}{\partial x}\left(2\mu\frac{\partial u}{\partial x}\right)+\frac{\partial}{\partial y}\left(\mu\left(\frac{\partial u}{\partial y}+\frac{\partial v}{\partial x}\right)\right)+\frac{\partial}{\partial z}\left(\mu\left(\frac{\partial u}{\partial z}+\frac{\partial w}{\partial x}\right)\right)$$
(17)

The viscous terms contain the first partial derivative of the velocity. Considering the effects of the eddy, they can be replaced by the actual first derivatives as listed in **Equation (10)** and similar expressions. Consequently, **Equation (17)** can be rewritten as follows:

$$\rho\left(\frac{\partial u}{\partial t}+u\frac{\partial u}{\partial x}+v\left(\frac{\partial u}{\partial y}+\delta\frac{\partial^2 u}{\partial y^2}\right)+w\left(\frac{\partial u}{\partial z}+\delta\frac{\partial^2 u}{\partial z^2}\right)\right)=$$
$$-\frac{\partial p}{\partial x}+\rho g_x+\frac{\partial}{\partial x}\left(2\mu\frac{\partial u}{\partial x}\right)+$$
$$\frac{\partial}{\partial y}\left(\mu\left(\left(\frac{\partial u}{\partial y}+\delta\frac{\partial^2 u}{\partial y^2}\right)+\left(\frac{\partial v}{\partial x}+\delta\frac{\partial^2 v}{\partial x^2}\right)\right)\right)+$$
$$\frac{\partial}{\partial z}\left(\mu\left(\left(\frac{\partial u}{\partial z}+\delta\frac{\partial^2 u}{\partial z^2}\right)+\left(\frac{\partial w}{\partial x}+\delta\frac{\partial^2 w}{\partial x^2}\right)\right)\right)$$
(18)

The continuity equation can be expressed as follows:

$$\frac{\partial u}{\partial x}+\frac{\partial v}{\partial y}+\frac{\partial w}{\partial z}=0 \tag{19}$$

By substituting **Equation (19)** into **Equation (18)**, the governing equation can be obtained by appropriate simplification and rearrangement, as follows:

$$\rho\left(\frac{\partial u}{\partial t}+u\frac{\partial u}{\partial x}+v\frac{\partial u}{\partial y}+w\frac{\partial u}{\partial z}\right)=-\frac{\partial p}{\partial x}+\rho g_x+\mu\left(\frac{\partial^2 u}{\partial x^2}+\frac{\partial^2 u}{\partial y^2}+\frac{\partial^2 u}{\partial z^2}\right)-$$
$$\delta\rho\left(v\frac{\partial^2 u}{\partial y^2}+w\frac{\partial^2 u}{\partial z^2}\right)+\delta\mu\left(-\frac{\partial^3 u}{\partial x^3}+\frac{\partial^3 u}{\partial y^3}+\frac{\partial^3 u}{\partial z^3}\right)$$
(20)

where,

$$R_1(u)=\delta\rho\left(v\frac{\partial^2 u}{\partial y^2}+w\frac{\partial^2 u}{\partial z^2}\right)$$
$$R_2(u)=\delta\mu\left(-\frac{\partial^3 u}{\partial x^3}+\frac{\partial^3 u}{\partial y^3}+\frac{\partial^3 u}{\partial z^3}\right)$$
(21)

Similar governing equations can be obtained in the *y*- and *z*-directions. The governing equations can then be rewritten in nondimensional forms, as follows:
(1) The continuity equation can be expressed as follows:

$$\frac{\partial u}{\partial x}+\frac{\partial v}{\partial y}+\frac{\partial w}{\partial z}=0 \tag{22}$$

(2) The momentum equations: *x*-axis direction:

$$\frac{\partial u}{\partial t}+u\frac{\partial u}{\partial x}+v\frac{\partial u}{\partial y}+w\frac{\partial u}{\partial z}=-\frac{\partial p}{\partial x}+g_x+\frac{1}{Re}\left(\frac{\partial^2 u}{\partial x^2}+\frac{\partial^2 u}{\partial y^2}+\frac{\partial^2 u}{\partial z^2}\right)-\frac{\delta}{L}\left(v\frac{\partial^2 u}{\partial y^2}+w\frac{\partial^2 u}{\partial z^2}\right)+\frac{\delta}{L}\frac{1}{Re}\left(-\frac{\partial^3 u}{\partial x^3}+\frac{\partial^3 u}{\partial y^3}+\frac{\partial^3 u}{\partial z^3}\right) \tag{23}$$

In the *y*-axis direction:

$$\frac{\partial v}{\partial t}+u\frac{\partial v}{\partial x}+v\frac{\partial v}{\partial y}+w\frac{\partial v}{\partial z}=-\frac{\partial p}{\partial y}+g_y+\frac{1}{Re}\left(\frac{\partial^2 v}{\partial x^2}+\frac{\partial^2 v}{\partial y^2}+\frac{\partial^2 v}{\partial z^2}\right)-\frac{\delta}{L}\left(u\frac{\partial^2 v}{\partial x^2}+w\frac{\partial^2 v}{\partial z^2}\right)+\frac{\delta}{L}\frac{1}{Re}\left(\frac{\partial^3 v}{\partial x^3}-\frac{\partial^3 v}{\partial y^3}+\frac{\partial^3 v}{\partial z^3}\right) \tag{24}$$

In the *z*-axis direction:

$$\frac{\partial w}{\partial t}+u\frac{\partial w}{\partial x}+v\frac{\partial w}{\partial y}+w\frac{\partial w}{\partial z}=-\frac{\partial p}{\partial z}+g_z+\frac{1}{Re}\left(\frac{\partial^2 w}{\partial x^2}+\frac{\partial^2 w}{\partial y^2}+\frac{\partial^2 w}{\partial z^2}\right)-\frac{\delta}{L}\left(u\frac{\partial^2 w}{\partial x^2}+v\frac{\partial^2 w}{\partial y^2}\right)+\frac{\delta}{L}\frac{1}{Re}\left(\frac{\partial^3 w}{\partial x^3}+\frac{\partial^3 w}{\partial y^3}-\frac{\partial^3 w}{\partial z^3}\right) \tag{25}$$

where $\delta$ denotes the scale coefficient, $L$ denotes the characteristic dimension of the solution domain, and $Re$ is the Reynolds number.

**Equations (22)–(25)** are extended N-S equations that can be solved using the appropriate initial and boundary conditions.

## 2.2 The properties of the extended equations

1) The extended N–S equations are compatible with the classical N–S equations, and by setting $\delta=0$, we can obtain the classical N–S equations. In general, the angular velocity of an eddy is not equal to the vorticity-induced angular velocity of the differential element, this difference in angular velocity changing the velocity distribution of the differential element, causing additional convection and diffusion. In the extended equations, $R_1(\boldsymbol{u})$ denotes the additional convection and $R_2(\boldsymbol{u})$ denotes the additional diffusion. It should be noted that the rotation of the eddy does not directly generate additional normal stress, the normal stress parts in the additional diffusion terms arising from the need to maintain the continuity of flow obtained by substituting the continuous equation. There is another explanation for obtaining extended N–S equations from the classical N‐S equations. Based on the previous hypothesis, the differential element is attached to the eddy, which is a local non-inertial rotating frame with an angular velocity, the magnitude of which is the difference in the angular velocity described above, the N–S equations being built on this frame. Transforming the N–S equations into the global coordinate system from this non-inertial frame, the extended N–S equations can be obtained, $R_1(\boldsymbol{u})$ can be interpreted as the Coriolis force, $R_2(\boldsymbol{u})$ being the additional diffusion term after the coordinate transformation [15].

2) The extended equations describe the laminar and turbulent flow in a unified manner, the additional terms of the extended equations representing the sources of velocity fluctuations. Eddies gain energy from the flow field, deviate from the original streamline, and then evolve over time to produce velocity fluctuations, further increasing the size and number of vortices, and forming velocity fluctuations over a wide range of time and length scales. When the Reynolds number is low, the rotating motion of the eddy does not have enough energy to sustain itself, and the fluctuations quickly dissipate into internal thermal energy, and the flow remains steady, this flow pattern being laminar flow; as the Reynolds number increases, the rotating motion has more energy, and the vortex flow intensifies, eventually forming turbulence, making the flow extremely complicated. In the flow calculation, when the fluid flow is laminar, the influence of additional terms is very small and can be omitted, whereas numerical calculations using the classical N–S equations can ensure sufficient accuracy; in the turbulent flow, the influence of velocity fluctuations is significant, and the additional terms must be retained to obtain the correct results. Moreover, the extended N–S equation can be implemented uniformly from laminar to turbulent flow without the need to distinguish between the different flow states.

3) In fluid flow simulations, the equations describing the mean flow quantities can be obtained directly from the extended N–S equations. In fluid flow, the eddies represent multi-scale structures in the overall flow, the characteristics of large-scale eddies depending strongly on the geometry of the flow boundaries. The characteristics of small-scale eddies, however, are generally less related to the flow boundaries, some characteristics being universal and scale-similar—that is, there exists a property in the N–S equations known as invariance of the scale transformation [16–17]. Based on these properties, in fluid flow simulations the solution domain can be decomposed into the resolved scale and the unresolved scale, the flow in the resolved scale including the mean flow and large eddies, and the small eddies in the unresolved scale representing the source of the fluctuations. This operation is similar to that of the LES; however, it does not require a filtering function. As the source of the fluctuations is explicitly included in the extended N–S equations, the mean flow equations can be obtained by simply replacing the instantaneous hydrodynamic quantities and scale coefficient with the mean quantities and scale coefficient at the resolved scale in **Equations (22)–(25)**, as follows:

Continuity equation:

$$\frac{\partial \overline{u}}{\partial \overline{x}} + \frac{\partial \overline{v}}{\partial \overline{y}} + \frac{\partial \overline{w}}{\partial \overline{z}} = 0 \tag{26}$$

Momentum equations: *x*-axis direction

$$\frac{\partial \overline{u}}{\partial \overline{t}} + \overline{u}\frac{\partial \overline{u}}{\partial \overline{x}} + \overline{v}\frac{\partial \overline{u}}{\partial \overline{y}} + \overline{w}\frac{\partial \overline{u}}{\partial \overline{z}} = -\frac{\partial \overline{p}}{\partial \overline{x}} + g_x + \frac{1}{\text{Re}}\left(\frac{\partial^2 \overline{u}}{\partial \overline{x}^2} + \frac{\partial^2 \overline{u}}{\partial \overline{y}^2} + \frac{\partial^2 \overline{u}}{\partial \overline{z}^2}\right) - \frac{\overline{\delta}}{L}\left(\overline{v}\frac{\partial^2 \overline{u}}{\partial \overline{y}^2} + \overline{w}\frac{\partial^2 \overline{u}}{\partial \overline{z}^2}\right) + \frac{\overline{\delta}}{L}\frac{1}{\text{Re}}\left(-\frac{\partial^3 \overline{u}}{\partial \overline{x}^3} + \frac{\partial^3 \overline{u}}{\partial \overline{y}^3} + \frac{\partial^3 \overline{u}}{\partial \overline{z}^3}\right) \quad (27)$$

where $\overline{u}$, $\overline{v}$, $\overline{w}$, and $\overline{p}$ denote the mean hydrodynamic quantities obtained in the flow simulation and $\overline{\delta}$ denotes the scale coefficient at the resolved scale.

It should be noted that the momentum equations in the *y*- and *z*-axes are similar to those in the *x*-axis.

In this section, we discuss a method for determining the scale coefficient at the resolved scale. As previously defined for the scale coefficient, the diameter of the eddy at the smallest scale is proportional to $v^{3/4}$, and the scale coefficient $\delta = (kv^{3/4})/4$ is only related to the kinematic viscosity of the fluid and is independent of the position of the eddy. With scale invariance in the N–S equations, the following transformation formula exists for viscosity at different scales:

$$v' \to \lambda^{1+H} v \quad (28)$$

where $\lambda$ denotes the scale factor and $H$ denotes a scaling exponent. For fully developed turbulence in the inertia region, $H = 1/3$.

This leads to the assumption that the equivalent kinematic viscosity of the fluid is $\lambda^{4/3} v$ at the resolved scale; here, $H = 1/3$, the equivalent eddy is the smallest eddy at the resolved scale, and its diameter is proportional to $\left(\lambda^{4/3} v\right)^{3/4}$, so $\overline{\delta} = (kv^{3/4}\lambda)/4$. The scale coefficient at the resolved scale depends only on the kinematic viscosity and scale and is independent of the position of the equivalent eddy. In the numerical simulations, it can be assumed that the characteristic dimension of the mesh is $\Delta l$. The diameter of the equivalent eddy is proportional to $v^{3/4}\Delta l$, and the scale coefficient at the resolved scale can be expressed as $\overline{\delta} = (kv^{3/4}\Delta l)/4$. The value of $kv^{3/4}$ can be determined based on numerical experience and experimental measurements [18].

## 3. Results and Discussion: Test problem for the extended N–S equations

Flow around obstacles is common in nature and in many applications. Consequently, we selected the flow over a cube placed in a channel as an example to verify the proposed model. This test case has several features—that is, it is geometrically simple, but exhibits many important complex flow features, as shown in **Figure 3**. This case was investigated experimentally in [19] and has become a popular test case for LES [20]. In this study, the LES and experimental results were compared.

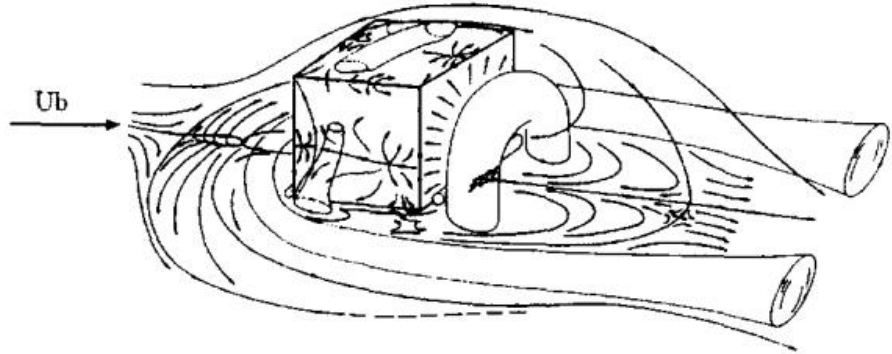

**Figure 3:** Schematic representation of the flow around a cube from [21].

The solution domain is shown in **Figure 4**. The height of the cube is the characteristic length (H), and H = 1 m. The length of the channel is 10·H, the width is 7·H, and the height of the channel is 2·H. The Reynolds number is defined based on the maximum inflow velocity and the height of the cube. The inflow velocity profile in the *x*-direction can be obtained by another channel simulation in which the Reynolds number is 3000 and the maximum inflow velocity is set to $U_b$ = 1 m/s [21]. A convective-type boundary condition can be applied at the outflow boundaries—that is, for the upper/lower boundaries and wall surfaces of the cube, no-slip conditions are used, assuming that both sides in the spanwise direction are symmetrical boundaries, and slip conditions are used on both sides. For comparison with existing results, a simulation was performed at a given Reynolds number, Re = 40,000.

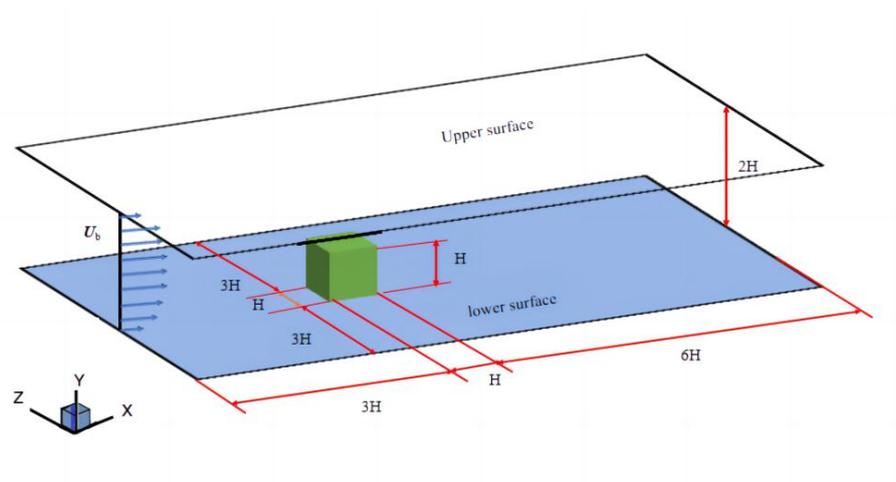

**Figure 4:** The solution domain for the flow over a cube placed in a channel.

The simulation was implemented using the conventional finite difference method with unknown variables located on staggered grids. The convective terms can be treated using the second-order SMART (Sharp And Monotonic Algorithm for Realistic Transport) upwind scheme, the viscous terms adopting the central difference scheme, and a second-order explicit Adams–Bashforth scheme being employed for the time discretization. Using a simple extension to approximate flows in arbitrary three-dimensional geometries, the cells of the solution domain can be divided into fluid and obstacle cells, the extended N–S equations being solved only in the fluid cells using the Chorin-type projection method. This algorithm is implemented in detail as described in Ref. [22].

The solution domain can be discretized using equidistant grids of $100 \times 40 \times 70$, the interior of the cube being approximated by marking it as a combination of obstacle cells. Based on the previous discussion, the characteristic dimension of the mesh can be considered to be the minimum value of the mesh in the three directions, and the scale coefficient ($\bar{\delta}$) can be obtained as follows:

$$\bar{\delta} = K \cdot \min(\Delta x, \Delta y, \Delta z)/4 \qquad (29)$$

where $K = k\nu^{3/4}$ is typically between 1 and 3; this value increasing appropriately with increasing Re [18].

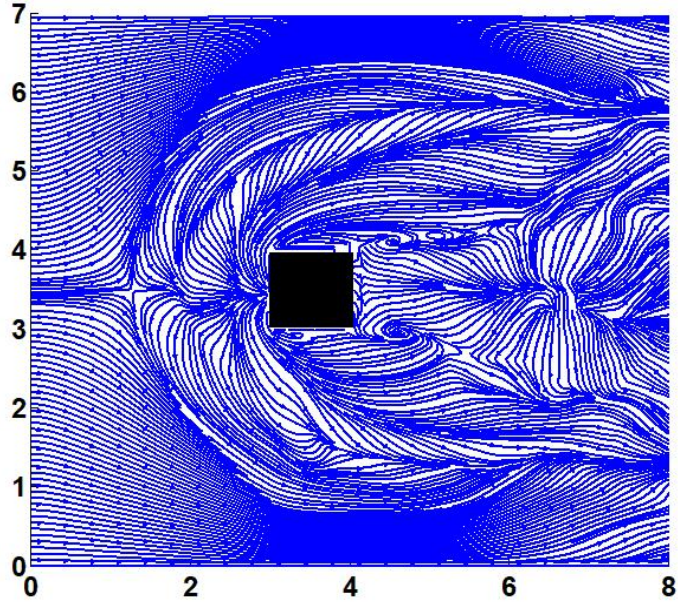

**Figure 5:** The time-averaged streamlines on the floor of the channel at y/H = 0.001.

For comparison with the experimental and LES results, three profiles were created at y/H = 0.001, z/H = 3.5, and x/H = 4.1, and averaged over a sufficiently long period. **Figure 5** shows the time-averaged velocity streamlines near the bottom surface of the plate at y/H = 0.001, corresponding to the flow visualization patterns obtained using oil streak techniques [19]. These streamlines reveal complex flow characteristics and contain significant information. The incoming flow reaches a saddle point in front of the body before separating around the body to form converging and diverging streamlines on the sides of the cube, thereby generating a horseshoe vortex. Two vortices are attached near the wall on both sides of the cube. Downstream, two footprints of the arch vortex are symmetrically distributed behind the cube, and there is a reattachment line further downstream of the cube. All the flow features in this simulation are consistent with the visualization results obtained using the oil streak techniques and the LES results.

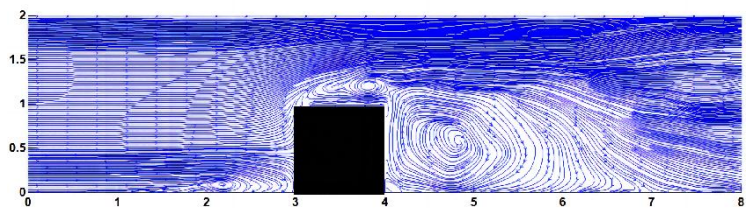

**Figure 6**: The time-averaged streamlines in the vertical center-plane of the flow over a cube at z/H = 3.5.

**Figure 6** shows the streamlines of the time-averaged flow in the profile at z/H = 3.5. Compared with the global flow characteristics shown in **Figure 3**, many of the flow characteristics are evident in **Figure 6**. At the front of the cube, there is a separation zone with a vortex at the bottom of the plate and the head of the horseshoe vortex. At the back of the cube, there is an arch vortex and reattachment line. At the top of the cube, a recirculation zone is attached to the top surface, forming a vortex near it.

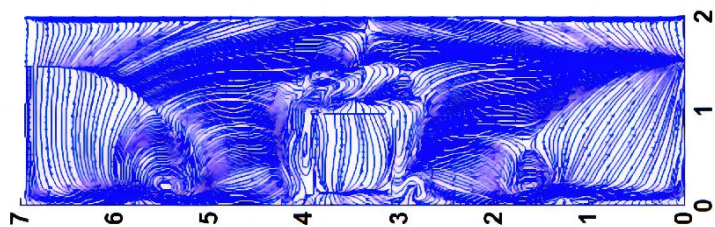

**Figure 7**: The time-averaged streamlines of the flow over a cube onto a plane parallel to the back face at x/H = 4.1.

    **Figure 7** shows the time-averaged streamlines on a plane parallel to the back of the cube at x/H = 4.1, where both horseshoe vortices are clearly visible. Compared to **Figure 3**, the time-averaged streamline plots of the three profiles are sufficient to represent the flow characteristics of the entire simulation object. These results are essentially in agreement with those of the experiment and the LESs.

    The time-averaged velocity distributions in the *y*-direction at x = 1.5H, z = 0, x = 2.5H, z = 0, x = 4.0H, and z = 0 are shown in **Figure 8**. The simulation results are compared with the LES results obtained using the Smagorinsky subgrid model and the measured velocity profiles; the LES results and experimental data being obtained from [8]. From **Figure 8**, it is evident that the simulation results agree with the results of the LES and measured data [7–8, 23–24].

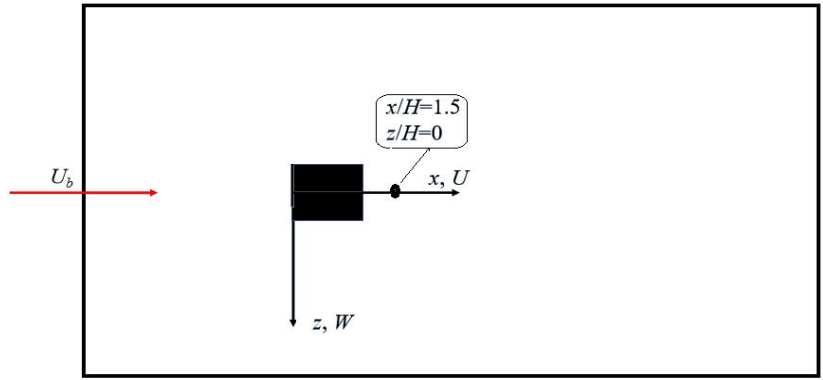

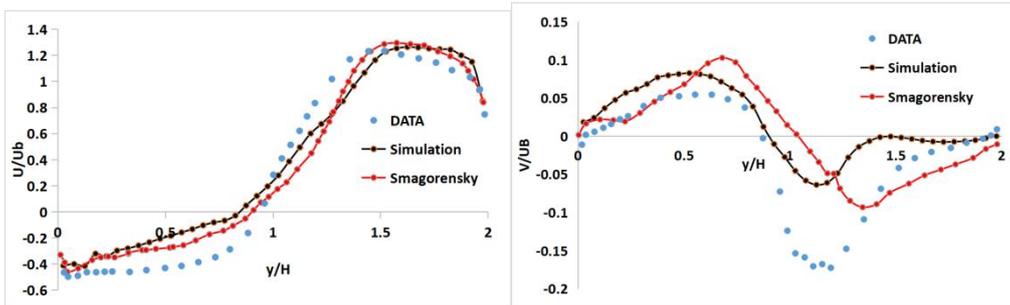

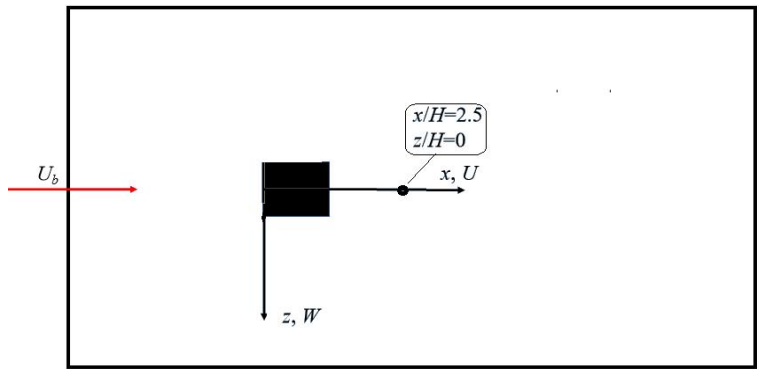

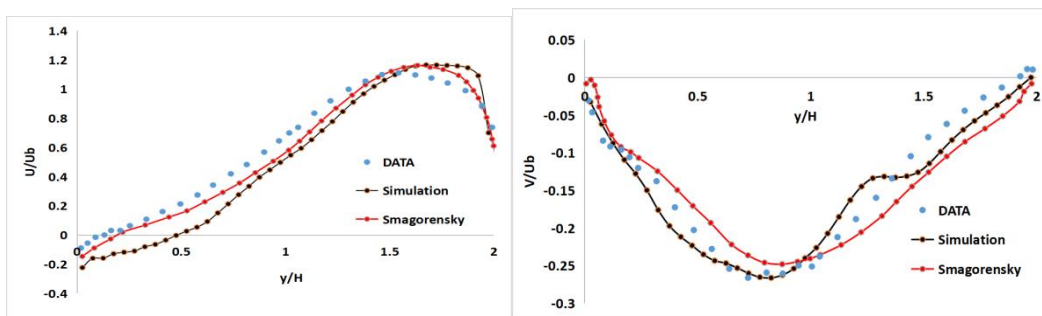

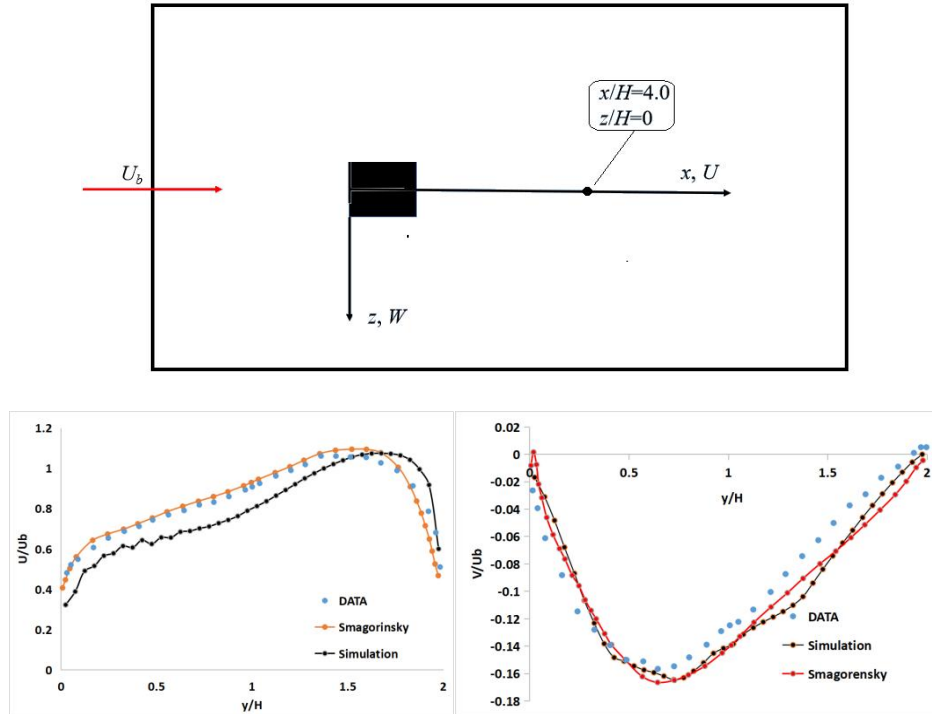

**Figure 8:** Results of the simulation, comparison with results of the LES and the experimental measurement.

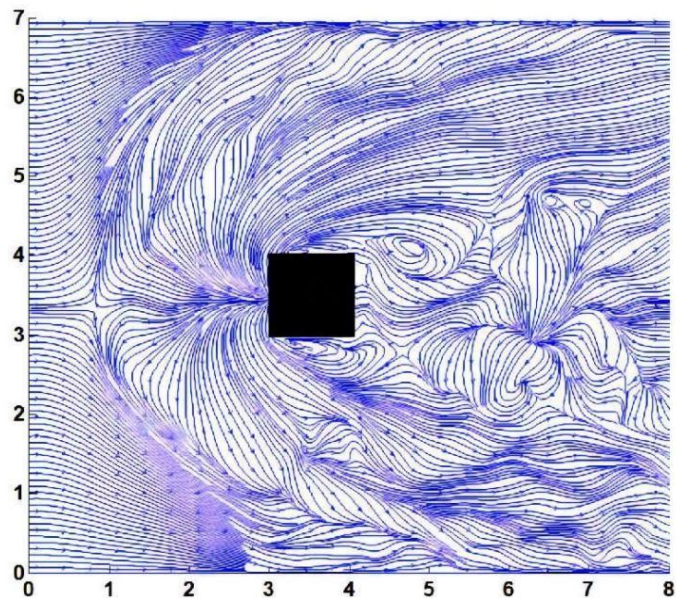

**Figure 9**: The instantaneous streamlines on the floor of the channel at y/H = 0.001.

It should be emphasized that instantaneous flow is very different from time-averaged flow. Comparing **Figures 5** and **9**, if we look only at the instantaneous flow, it is evident that the arch vortices in the flow are always asymmetrically distributed behind the cube; when the average time is sufficiently long, near-symmetrical arch vortices are visible.

## 4. Conclusions

The simulation of turbulent flow is a major challenge in computational fluid dynamics (CFD). The DNS of turbulent flow is unfeasible for engineering applications in the foreseeable future; for RANS and LES, the closure of the governing equations depends on the appropriate turbulence model; however, there have been no general-purpose turbulence closure models until now, the main problem being that the mean flow equations cannot be directly obtained from the N–S equations in turbulence simulations. In this study, through phenomenological observations and Kreuer's interpretation of the origin of viscosity, a basic hypothesis was established, based on which classical N–S equations were improved to obtain extended N–S equations. The extended N–S equations have a clear physical meaning, and the newly introduced scale coefficient has a clear meaning and is easy to determine. In the extended N–S equations, the equations are closed, and the source of the velocity fluctuations is explicitly considered; consequently, we can obtain the equations describing the average quantities directly from the extended equations without any additional turbulence model in the turbulence simulation. Finally, the flow over a cube placed in a channel was numerically investigated to verify the extended equations, the simulation results agreeing well with the LES and experimental results. The following conclusions could be drawn:

1) The extended N–S equations are compatible with the classical N–S equations. The extended equations were set up based on a hypothesis, which was based on phenomenological observations and Kreuer's interpretation of the origin of viscosity, and we believe this hypothesis to be rational. The extended N–S equations can be understood as the result of coordinate transformation of the N–S equations, the N–S equations being built on a local non-inertial rotating frame, transforming the N–S equations into the global coordinate system from this local frame, to obtain the extended N–S equations.

2) The source of the velocity fluctuations was explicitly included in the extended N–S equations. In general, the angular velocity of the eddy is not equal to the vorticity-induced angular velocity of the differential element, and this difference in angular velocity is the source of velocity fluctuations. We argued that the fluctuations are intrinsic to the fluid flow. In laminar flow, the velocity fluctuations do not have enough energy to sustain themselves and quickly dissipate into internal thermal energy, and the flow remains steady, whereas in turbulent flow, the velocity fluctuations have more energy and are constantly intensified, eventually forming turbulence. Thus, the extended equations can describe laminar and turbulent flows in a unified manner.

Because this is a new calculation method, the next step would be to implement it, using the finite-volume method or other numerical methods. Moreover, the method must be verified using a large number of actual flow problems to validate the new model.

**Acknowledgments**

This study was supported by the National Natural Science Foundation of China (Grant No. 52079118), the Guang'An Science and Technology Bureau Key Research and Development Plan (2022GYF01), and the Innovation Star Project of Gansu Province (2023CXZX-438).